\documentclass[manuscript]{acmart}
\usepackage{subcaption}
\usepackage{amsfonts}
\usepackage{tabularx}
\usepackage{microtype}
\usepackage{stfloats}
\usepackage{balance}
\usepackage{xcolor}
\usepackage{float}
\usepackage{multirow}
\usepackage{graphicx}

\newcommand{\eg}{\textit{e.g.}, }

\AtBeginDocument{%
  \providecommand\BibTeX{{%
    \normalfont B\kern-0.5em{\scshape i\kern-0.25em b}\kern-0.8em\TeX}}}

\setcopyright{none}

\begin{document}


\title{Distributed Cognition for AI-supported Remote Operations: Challenges~and~Research~Directions}

\begingroup
\renewcommand\thefootnote{}\footnote{
 This paper was presented at the 2025 ACM Workshop on Human-AI Interaction for Augmented Reasoning (AIREASONING-2025-01). This is the authors’ version for arXiv.}
\endgroup


\author{Rune M. Jacobsen}
\email{runemj@cs.aau.dk}
\orcid{0000-0002-1877-1845}
\affiliation{%
  \institution{Aalborg University}
  \city{Aalborg}
  \country{Denmark}
}

\author{Joel Wester}
\email{joelw@cs.aau.dk}
\orcid{0000-0001-6332-9493}
\affiliation{%
  \institution{Aalborg University}
  \city{Aalborg}
  \country{Denmark}
}

\author{Helena Bøjer Djernæs}
\email{hbd@cs.aau.dk}
\orcid{0009-0003-9457-811X}
\affiliation{%
  \institution{Aalborg University}
  \city{Aalborg}
  \country{Denmark}
}

\author{Niels van Berkel}
\email{nielsvanberkel@cs.aau.dk}
\orcid{0000-0001-5106-7692}
\affiliation{%
  \institution{Aalborg University}
  \city{Aalborg}
  \country{Denmark}
}

\renewcommand{\shortauthors}{Jacobsen et al.}

\begin{abstract}
This paper investigates the impact of artificial intelligence integration on remote operations, emphasising its influence on both distributed and team cognition. As remote operations increasingly rely on digital interfaces, sensors, and networked communication, AI-driven systems transform decision-making processes across domains such as air traffic control, industrial automation, and intelligent ports. However, the integration of AI introduces significant challenges, including the reconfiguration of human-AI team cognition, the need for adaptive AI memory that aligns with human distributed cognition, and the design of AI fallback operators to maintain continuity during communication disruptions. Drawing on theories of distributed and team cognition, we analyse how cognitive overload, loss of situational awareness, and impaired team coordination may arise in AI-supported environments. Based on real-world intelligent port scenarios, we propose research directions that aim to safeguard human reasoning and enhance collaborative decision-making in AI-augmented remote operations. 
\end{abstract}





\maketitle
\balance

\section{Introduction}
The introduction of artificial intelligence (AI) into remote operations will have a tremendous effect on operators' cognition~\cite{Zhang2019, Veitch2022}. Remote operation refers to the process of monitoring, controlling, or supervising systems, machinery, or environments from a distance using remote user interfaces~\cite{Gafert2022, Tener2022}, sensors, and networked communication technologies~\cite{Sanders2006}. These operations depend on real-time information visualisation to support operators in complex decision-making scenarios. Remote operations are widely used in domains where direct physical presence is impractical, costly, or hazardous~\cite{Cramer2011}. Examples include air traffic control~\cite{Gawade2016}, industrial automation~\cite{DeAlwis2024}, and critical infrastructure management. While AI-driven solutions can introduce efficiency gains in these domains, there may also be unintended consequences. Here, an increased reliance on AI could have profound implications for the operators' cognitive functions, for example how the operators distribute cognition for themselves~\cite{Hollan2000, hutchins1995cognition} and how it affects team cognition for operators~\cite{MacMillan2004}. 

Contemporary remote work and teleoperation scenarios often exhibit distributed cognition, where cognitive processes are shared across human operators and tools in the environment~\cite{Hollan2000}, and now also AI systems. At the same time, team cognition, described as shared knowledge, mental models and information processing for team coordination, emerges as a critical factor. Put differently, a team’s collective cognitive state is not just the sum of individuals, but a product of communication and coordination that yields a shared understanding of tasks and situations~\cite{Canonico2019, wester2024theory}. However, introducing AI into these teams in remote operations also creates challenges. Prior work has, for example,identified risks of cognitive overload, as humans must manage complex AI-driven information and alerts, and loss of situational awareness when automation handles tasks out-of-sight~\cite{Rebensky2022}. In turn, team dynamics can suffer when AI agents act without (effectively) communicating their intentions, human teammates may consequently struggle to maintain common ground, leading to worsened coordination~\cite{Liang2019}. These challenges call for reimagining cognitive support systems with solutions that not only integrate AI into augmented workflows but also actively protect human situational awareness and team cohesion, transforming remote operations into environments where technology and human expertise work in unison.

In this paper, we highlight the impact of AI integration in remote operations and the challenges it presents to distributed cognition among operators and team cognition in human-AI collaboration. Grounded in the theories of distributed and team cognition, we identify three key areas of concern that arise in AI-supported remote operations: (1) reconfiguring team cognition in human-AI teams, (2) adapting AI memory to align with human distributed cognition, and (3) leveraging AI as a fallback operator during communication disruptions. For each area of concern we describe pertinent challenges that emerges. To illustrate these challenges in a real-world setting, we focus on intelligent ports, demonstrating how AI-driven decision-making may affect coordination, memory structures, and resilience in remote environments. Based on these insights, we propose concrete research directions aimed at supporting and safeguarding human reasoning in AI-augmented workflows, ensuring that AI enhances, rather than disrupts, cognitive processes in remote operations.

\section{Related Work}
Distributed cognition is a theoretical framework that extends the traditional view of cognition beyond an individual’s mind to include interactions between people, artifacts, and their environment. Hutchins~\cite{hutchins1995cognition, Hutchins1995cockpit} introduced this concept through studies of ship navigation and cockpit operations, arguing that cognition is distributed across multiple memory and processing units, including both human and non-human components. In these settings, knowledge is not stored solely within individuals but is externalised in artifacts such as maps, logbooks, and instruments, which enable coordination and decision-making~\cite{rogers2005cognition}. This perspective shifts the focus from isolated cognitive processes to a broader system of interdependent elements working together to accomplish goals, such as steering a ship or flying an aircraft~\cite{Hutchins1995cockpit}. DC highlights three core aspects of distributed information processing: how knowledge is represented and shared, how communication occurs as an activity unfolds, and how coordination is achieved across distributed units~\cite{perry2003distributed}. For example, in Hutchins’~\cite{hutchins1995cognition} analysis of ship navigation, he describes how one worker records positional information in a logbook, which is then interpreted by a team member who recalibrates the navigational tools accordingly. The coordination of these individual contributions allows the system as a whole to function effectively. By framing cognition as a socio-technical process embedded in dynamic interactions, DC provides a lens to analyse how individuals, teams, and artifacts collectively solve complex problems.

In collaborative environments, a term often used is team cognition~\cite{McNeese2020}, which examines how groups of individuals collaboratively manage and process information to achieve shared objectives~\cite{MacMillan2004}. Teams operating in dynamic, high-stakes settings, such as emergency response, aviation, or remote operations, rely on effective information flow, coordination, and shared mental models to function efficiently~\cite{MacMillan2004, Entin1999}. A key concept in team cognition is implicit coordination, where team members anticipate each other’s needs and actions without the need for explicit communication~\cite{MacMillan2004, Serfaty1993}. This anticipation reduces communication overhead, allowing team members to focus on task execution rather than constantly negotiating roles and responsibilities~\cite{MacMillan2004}. Implicit coordination is particularly beneficial in time-sensitive environments, where excessive communication can increase cognitive load and slow down decision-making~\cite{MacMillan2004}. Shared mental models play a crucial role in this process by enabling team members to develop a mutual understanding of both their own tasks, roles, and expected behaviours~\cite{Entin1999}, as well as the tasks of others in a team. These models help individuals predict the actions of their teammates and adjust their behaviour accordingly, enhancing synchronisation and reducing misunderstandings~\cite{Endsley1995}. Situational awareness further supports team cognition by ensuring that members remain informed about the evolving context of their work, allowing them to react appropriately to unexpected changes~\cite{Endsley1995, Entin1999}. In distributed teams, particularly in remote or virtual operations, maintaining these cognitive mechanisms becomes more challenging due to the reliance on mediated communication and asynchronous collaboration. Nevertheless, applying a distributed cognition perspective to team-based work allows us to consider how cognitive processes extend beyond individual actors to encompass collective, technology-mediated workflows.

\section{Areas of Concern in AI-supported Remote Operations}
AI is rapidly taking on a growing role as decision-making or decision-support tool, shaping workflows, adapting to change, and operating with decreasing human intervention in remote operations. This shift challenges traditional models of team cognition, raising critical questions about how operators can interact with AI in remote operations. For both distributed cognition and team cognition, AI (with varying levels of autonomy) will affect how individuals and teams make decisions, how memory distributed across multiple entities may influence operations, as well as how decisions made during fallouts of communication can affect the distributed cognition. Using intelligent ports as a case study, the following sections explore key HCI challenges and research directions for designing AI that integrates into both individual and team-based decision making.

\subsection{Reconfiguring Human-AI Team Cognition}
Remote operations rely on distributed cognition, where human teams coordinate across distances using digital communication tools~\cite{Xisong2013, Wu2013}, automated systems~\cite{Yao2021}, and AI-supported decision making~\cite{Lam2012}. Traditionally, AI has been treated as a tool that aids human decision-making, optimising processes under human supervision. However, as remote operations become increasingly complex, AI is beginning to act as a more autonomous agent, making decisions in environments where immediate human oversight is not always feasible. This challenges traditional models of team cognition, which assume that humans are the primary agents responsible for coordination and decision making. When AI systems operate independently in remote environments, new interactional challenges emerge.

Intelligent ports exemplify these challenges, as they rely on AI-driven scheduling and planning systems to manage vessel traffic, cargo logistics, and predictive maintenance~\cite{Tang2019}. Unlike traditional ports, where human operators directly oversee operations, intelligent ports function with a mix of human supervision and semi-autonomous AI decision-making~\cite{DeAlwis2024}. When congestion occurs (\eg multiple vessels arriving simultaneously due to weather-related delays) various subsystems must collaboratively adjust docking schedules, reroute ships, and allocate resources. However, if these agents operate as isolated decision-makers without a shared interaction framework, human operators may struggle to maintain awareness of AI-driven decisions or to intervene effectively when adjustments are needed~\cite{jacobsen2020}. Without clear communication mechanisms between AI and human supervisors, operators may be forced to override AI decisions based on incomplete understanding.
To address this, HCI research must rethink how human-AI teams establish and maintain cognitive alignment. 

First, research should explore how AI can develop a communicative presence, not just through status updates, but through proactive engagement, feedback requests, and real-time collaborative problem-solving. This could involve AI using conversational interfaces, ambient notifications, or even spatialised audio cues (as envisioned in homes~\cite{Jacobsen2024}) to provide reasoning in an unobtrusive yet effective way. 

Second, current AI explainability research often focuses on post-hoc justifications, where AI explains why it made a decision after the fact. However, in remote operations, human operators should not only interpret AI’s past decisions but also be aware and potentially negotiate its future ones. We envision that instead of AI simply assigning a docking schedule in an intelligent port, it could engage in a lightweight, constraint-based negotiation with human operators---adapting its plan based on the operators' expertise and real-time situational awareness.

\subsection{Adapting AI Memory}
Remote operations generate vast amounts of data, requiring teams to filter, prioritise, and adapt information to changing circumstances~\cite{Wimber2015, DeJong2024} (\eg adaptive forgetting). Unlike humans, who naturally forget irrelevant details and adjust their focus based on context, AI systems typically retain data, potentially treating pieces of information as important~\cite{kaushik2021understanding} (\eg catastrophic remembering). While comprehensive memory may seem advantageous, it can lead to cognitive overload for human operators, outdated decision-making, and inefficiencies when AI retrieves or acts on obsolete information. In dynamic environments where remote operators must respond quickly to evolving conditions, AI’s inability to ``forget'' can become a liability rather than a strength. The challenge, therefore, is not just about memory storage but about how AI dynamically adapts what it remembers, forgets, and prioritises to maintain effective collaboration in remote teams.

Intelligent ports exemplify this issue, as they rely on AI-driven monitoring and coordination to optimise logistics, equipment usage, and vessel movements. Over time, these AI systems accumulate extensive records of ship docking schedules, crane assignments, and cargo movements~\cite{Valero2021}. While historical data is valuable for trend analysis and predictive maintenance, this kind of rigid memory retention can also create inefficiencies in real-time decision-making. For instance, if an intelligent port’s AI continues prioritising a crane based on its past reliability without factoring in recent wear-and-tear or maintenance logs, it may misallocate resources, leading to delays and inefficiencies. Without a mechanism for adaptive forgetting~\cite{Wimber2015, kaushik2021understanding}, AI systems risk overwhelming remote human operators with excessive alerts, unwarranted recommendations, or constraints that no longer reflect the current reality in the remote operations. 
To address this, HCI research must approach AI memory as interactive, adaptive, and explainable. 

First, it can be difficult for human operators to understand why AI retrieves certain information or how past decisions shape future recommendations. Future collaborative tools could provide interactive interfaces for the AI's memory, where users can view, query, edit, or even challenge AI’s recall~\cite{berkel2024}. 

Second, unlike humans, who naturally filter and forget information based on relevance, AI needs explicit mechanisms to prioritise what should be retained versus discarded. HCI research should investigate adaptive mechanisms, where AI dynamically de-prioritise, delays, or removes information based on operational context and human feedback. For instance, could remote operators receive \textit{``forgetting suggestions''} from the AI, prompting them to decide whether certain past rules or patterns should still influence future actions?

Third, memory is not just an AI challenge, it is a shared cognitive resource. Research should explore human-AI collaborative memory spaces, where both human operators and AI can jointly build, refine, and adapt a shared operational history. For instance, in intelligent ports, AI could maintain a dynamic knowledge graph of past disruptions, interventions, and operator adjustments, allowing humans to annotate, revise, or tag information for future AI reference.

\subsection{AI as a Fallback Operator}
Remote operations are highly dependent on stable digital infrastructure to maintain coordination between human teams and AI-driven systems~\cite{lind2015port}. However, network failures, cyberattacks, extreme weather conditions, or unforeseen technical malfunctions can disrupt real-time communication, leaving remote operators unable to intervene in critical situations~\cite{Hou2023, Gao2023}. Many AI systems have an assistive nature, meaning they either halt operations, wait for human re-engagement, or rigidly follow pre-programmed protocols when a disruption occurs. This creates a major operational risk as when communication is lost---who or what ensures continuity? In high-stakes remote environments, a lack of adaptable AI agency can lead to cascading failures, as AI remains passive in moments when rapid, autonomous decision-making is required. How to design AI that can temporarily assume responsibility, ensuring operational continuity during communication disruptions while facilitating a smooth transition back to human control once connectivity is restored is therefore a timely challenge. 

Intelligent ports highlight this challenge, as they rely on AI-driven logistics systems to manage vessel docking, cargo handling, and supply chain synchronisation. These ports depend on a complex web of interconnected intelligent systems and remote human oversight, making them vulnerable to failures in networked coordination. Imagine a scenario where multiple container ships are en route to an intelligent port when a connectivity failure isolates the port’s AI from remote human supervisors. If AI lacks fallback autonomy, it may either pause vessel docking, causing congestion and fuel waste, or rigidly follow pre-set schedules without accounting for new variables, such as worsening weather or an emergency requiring priority docking. Conversely, an AI system capable of temporarily assuming control could reconfigure docking priorities, reroute vessels, or coordinate local autonomous systems while signalling uncertainty levels and awaiting human re-engagement. 

First, the shift between human-led and AI-led operations should not be considered as a binary switch but a gradual, interactive transition. Research should explore not only how AI may signal autonomy escalation (\eg when it detects a disruption and takes over), but also how it prepares humans for re-engagement when connectivity is restored. 

Second, if AI must make decisions in the absence of human input, it must also communicate how certain or uncertain it is about those decisions after reconnecting. Research could explore how AI expresses its own confidence levels, flagging moments when its fallback autonomy is operating with incomplete or ambiguous data. For example, in an intelligent port AI might autonomously reroute a vessel but signal uncertainty due to missing weather reports, prompting a human operator to validate the decision once reconnected. 

Third, continuing in a similar vein, once human operators regain control, understanding what the system did in the user's absence is critical in restoring situational awareness. Research should investigate how AI can generate interpretable, interactive autonomy logs, where operators can review, adjust, or override AI-driven decisions.

\section{Conclusion}
In conclusion, integrating artificial intelligence into remote operations is not merely a technological upgrade, it is a fundamental shift that redefines how cognitive processes are distributed among human operators, AI systems, and collaborative teams. Our discussion has highlighted three critical areas of concern: the need to reconfigure human-AI team cognition, the challenges of adapting AI memory to dynamic operational contexts, and the importance of designing AI to function as a reliable fallback operator during communication disruptions. Using intelligent ports as a case study, we illustrated how these challenges manifest in real-world settings, where the potential benefits of AI-driven decision-making must be balanced against risks to situational awareness and team coordination. Moving forward, research must focus on creating systems and interfaces that foster transparent, adaptive, and resilient human-AI collaboration. By doing so, we can harness the efficiency and decision-making powers of AI while safeguarding the cognitive strengths and coordination of human teams. Ultimately, the goal is to ensure that as AI becomes an ever-more integral part of remote operations, it enhances rather than undermines the distributed and collective intelligence that lies at the heart of effective, safe, and complex operational practices.
\begin{acks}
This work was supported by a research grant (VIL69118) from Villum Fonden.
\end{acks}

\bibliographystyle{ACM-Reference-Format}
\bibliography{11-references}

\appendix

\end{document}